\documentclass[submission, Proceedings]{SciPost}

\hypersetup{
    colorlinks,
    linkcolor={red!50!black},
    citecolor={blue!50!black},
    urlcolor={blue!80!black}
}

\usepackage[bitstream-charter]{mathdesign}
\DeclareSymbolFont{usualmathcal}{OMS}{cmsy}{m}{n}
\DeclareSymbolFontAlphabet{\mathcal}{usualmathcal}

\begin{document}

\begin{center}{\Large \textbf{
Latest ALICE results on J/$\mathbf{\psi}$ photoproduction in ultra-peripheral collisions at the LHC\\
}}\end{center}

\begin{center}
T. Herman\textsuperscript{1$\star$} for the ALICE Collaboration
\end{center}

\begin{center}
{\bf 1} Faculty of Nuclear Sciences and Physical Engineering, Czech Technical University in Prague, Prague, Czech Republic
\\
* tomas.herman@cern.ch
\end{center}

\begin{center}
\today
\end{center}


\definecolor{palegray}{gray}{0.95}
\begin{center}
\colorbox{palegray}{
  \begin{tabular}{rr}
  \begin{minipage}{0.1\textwidth}
    \includegraphics[width=22mm]{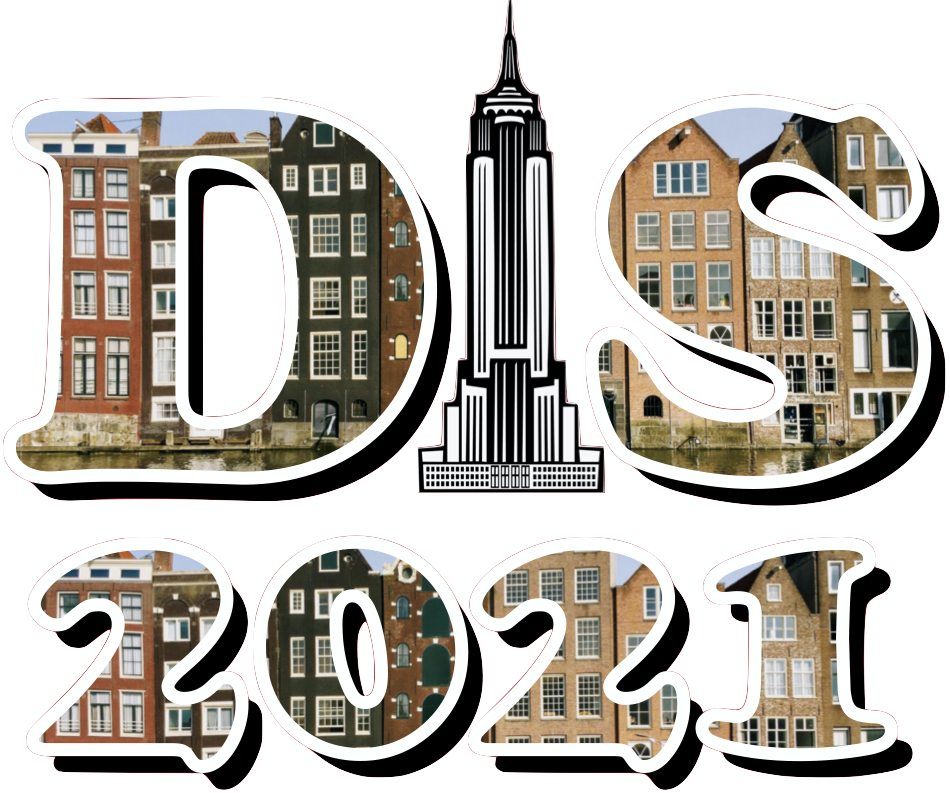}
  \end{minipage}
  &
  \begin{minipage}{0.75\textwidth}
    \begin{center}
    {\it Proceedings for the XXVIII International Workshop\\ on Deep-Inelastic Scattering and
Related Subjects,}\\
    {\it Stony Brook University, New York, USA, 12-16 April 2021} \\
    \doi{10.21468/SciPostPhysProc.?}\\
    \end{center}
  \end{minipage}
\end{tabular}
}
\end{center}

\section*{Abstract}
{\bf

Ultra-peripheral collisions (UPCs), studied using the ALICE detector, allow us to investigate the low-$\mathbf{x}$ behavior of the gluon distribution of the colliding particles. 

Two new measurements of coherent J/$\mathbf{\psi}$ photoproduction cross section from Pb–Pb UPCs at $\mathbf{\sqrt{s_{\rm NN}}}=5.02$ TeV are presented: the first measurement of the $\mathbf{|t|}$-dependence of the cross section providing a new tool to investigate the transverse gluonic structure at low Bjorken-$\mathbf{x}$ and a rapidity-differential measurement at midrapidity allowing us to provide stringent constraints on nuclear gluon shadowing and saturation models.

In addition, prospects for heavy vector meson photoproduction measurements in LHC Run 3 and 4 are presented.
}


\section{Introduction}
\label{sec:intro}
At low Bjorken-$x$ the gluon content of free nucleons or nucleons within nuclei is predicted to reach a saturation regime \cite{Mueller:1989st}, where the gluon distribution is expected to stop growing, which has not been conclusively observed yet. So far, there have been only hints of possible onset of saturation. Ultra-peripheral collisions allow us to investigate, using photon-induced interactions, processes sensitive to the low-$x$ behavior of the gluon distribution and they are a promising tool to provide important constraints on the initial stages of the collision. With increasing atomic number $A$ of the studied nucleus, there are stronger nuclear shadowing effects on the gluon PDFs at low $x$. The onset of saturation is expected to depend on $A$ and it may by one of the contributing factors to nuclear shadowing.

In UPCs, the impact parameter $b$ of the colliding particles is larger than the sum of their radii $R$. Hence the short-range hadronic interactions are suppressed. This enables photon-induced reactions to be measured. It should be noted that photon-induced reactions also contribute at $b < R_1+R_2$ as demonstrated by the $\rm{J/\psi}$ excess at very low $p_{\rm T}$ reported by ALICE \cite{jpsiexcess} or by the dilepton excess at low $p_{\rm T}$ by STAR \cite{dileptonexcess}.

Coherent photoproduction of $\rm{J/\psi}$ at the LHC is measured at the highest available energies with the Bjorken-$x$ of the probed gluon distribution depending on parameters of the collision as
\begin{equation}
x = \frac{M_{\rm{J/\psi}}}{\sqrt{s_{\rm NN}}}e^{\pm y},
\end{equation}
where $y$ is the rapidity of the produced vector meson with mass $M_{\rm{J/\psi}}$ coming from a collision with given center of mass energy per nucleon pair $\sqrt{s_{\rm NN}}$. Thus by measuring the cross section at different rapidities the gluon distribution at different Bjorken-$x$ can be explored. To study also the transverse-plane distribution of the gluons in the colliding particles one can look at the $|t|~(\sim p^2_{\rm T} )$ dependence of the cross section which is related by a 2D Fourier transform to the impact parameter.

\section{ALICE detector}
\label{sec:detector}

The silicon Inner Tracking System (ITS) is used to trigger events and measure particle tracks in the central rapidity region of the ALICE detector nearest to the beam pipe. To identify particles and measure tracks further from the interaction point, a Time Projection Chamber (TPC) consisting of a drift volume with multi-wire proportional chambers is used. To improve particle identification and assist in triggering, a multigap resistive plate chamber Time-of-Flight (TOF) detector is used as well. These detectors cover the midrapidity region and are enclosed in a solenoid magnet with magnetic field intensity $B = 0.5$ T. 

In the forward rapidity region there is an absorber stopping the majority of the produced charged particles except muons. These are then measured using five planes of a Muon Tracker with cathode pad chambers. The third plane is enclosed in a dipole magnet used for the momentum measurement. For triggering there is a Muon Trigger with resistive plate chambers protected by an additional Muon Wall absorber.

To enforce the exclusivity condition implicit in coherent photoproduction, the following detectors covering forward rapidities are used to impose activity vetoes: the V0 scintillator counter and the ALICE Diffractive (AD) scintillator counter. In addition, the V0 detector is used in luminosity measurement.

\section{Results}
\label{sec:results}

\subsection{Rapidity differential cross section}
\begin{figure}[h!t]
\centering
\includegraphics[width=0.5\textwidth]{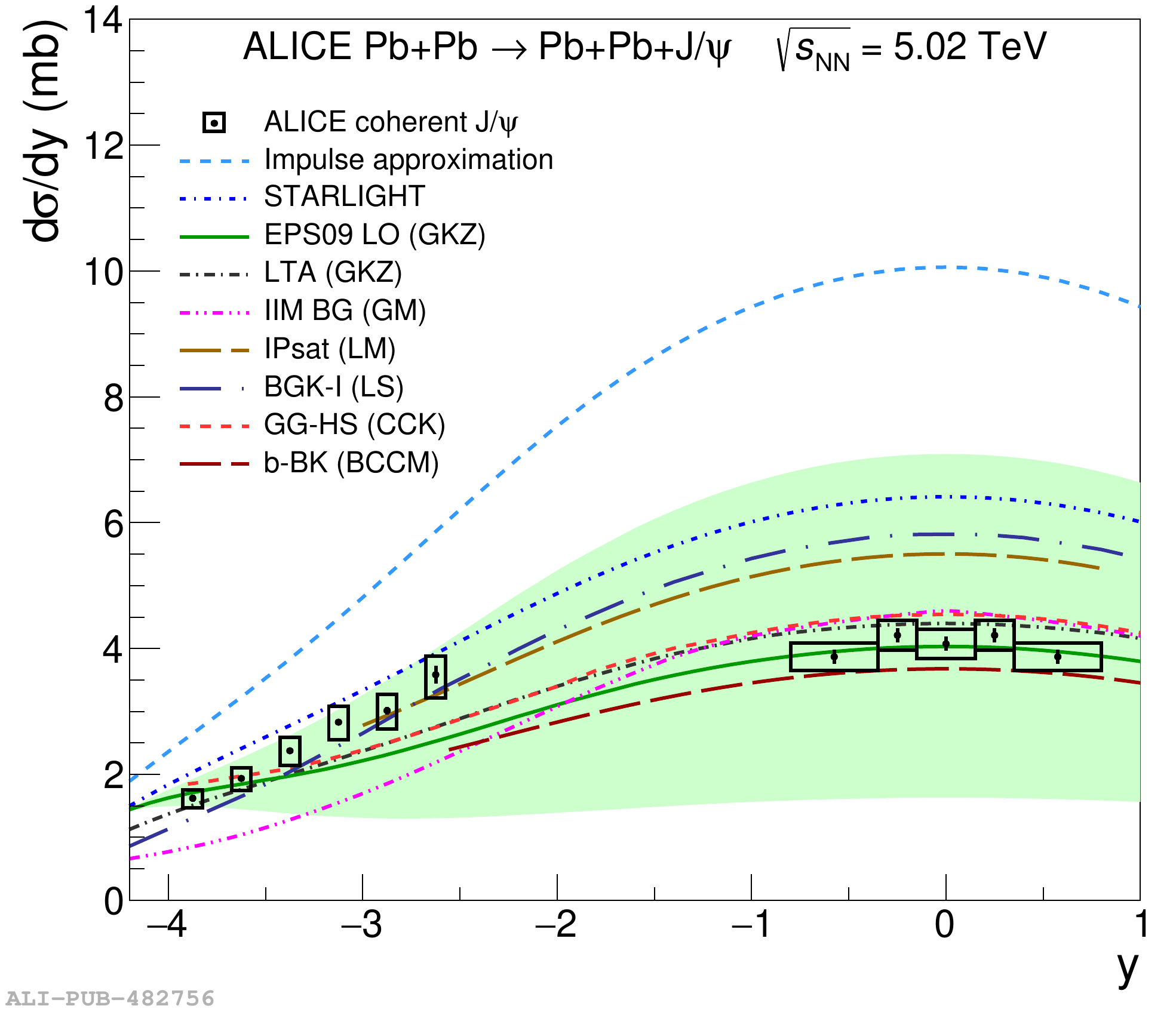}\includegraphics[width=0.5\textwidth]{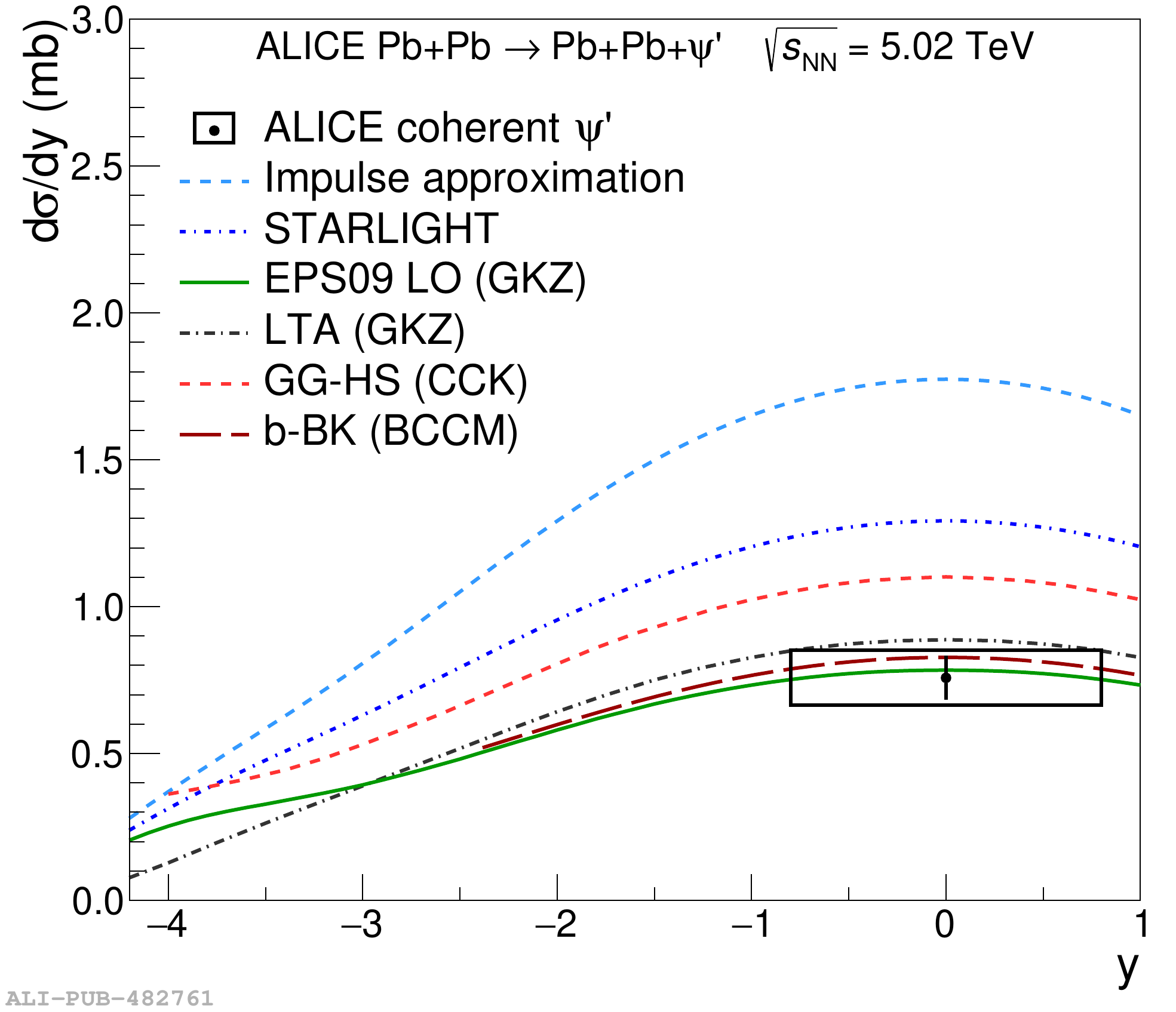}
\caption{Rapidity differential cross section of the coherent $\rm{J/\psi}$ (left) and $\psi'$ (right) photoproduction in Pb–Pb UPC events measured by ALICE. The error bars (boxes) show the statistical (systematic) uncertainties. Several theoretical models are also shown \cite{ypaper}.}
\label{fig:JPsi_y}
\end{figure}

The studied data sample consists of Pb–Pb UPCs at $\sqrt{s_{\rm NN}}=5.02$ TeV. The coherent $\rm{J/\psi}$ cross section at midrapidity is computed as a weighted average of the results obtained considering three decay channels: $\mu^+\mu^-, e^+e^-, p\bar{p}$. The new midrapidity measurement of the cross section \cite{ypaper} can be seen in Fig. \ref{fig:JPsi_y} (left) complementing the previously published forward rapidity values \cite{ALICE:2019tqa}. 

The ALICE results are compared to several theoretical predictions. First is the Impulse approximation (IA) based on photoproduction data from protons not including nuclear effects except coherence. Then there is the STARlight model also based on photoproduction data from protons but incorporating a vector meson dominance model to include multiple scattering, however without gluon shadowing. Next are the EPS09 LO prediction, which is a parametrization of nuclear shadowing data, and the Leading Twist Approximation (LTA) of nuclear shadowing. There is also a group of models based on the color dipole approach coupled to the Color Glass Condensate (CGC) formalism with different assumptions on the dipole-proton scattering amplitude (IIM BG, IPsat, BGK-I). Further there is the GG-HS prediction which is a color dipole model with hot-spot nucleon structure. And finally the b-BK model utilizing solutions of the impact-parameter dependent Balitsky-Kovchegov equation.

By taking the ratio of the measured cross section and the IA prediction the nuclear suppression factor $S_{\rm Pb}$ can be extracted for $x \in (0.3,1.4)\times 10^{-3}$ as
\begin{equation}
S_{\rm Pb} = \sqrt{\left. \left( \frac{{\rm d} \sigma}{{\rm d} y}\right)_{\rm data} \middle/ \left( \frac{{\rm d} \sigma}{{\rm d} y}\right)_{\rm IA} \right.} = 0.65 \pm 0.03.
\end{equation}
Models with shadowing (EPS09, LTA) and saturation (GG-HS) describe central and forward data but underestimate semi-forward data. Other models describe either the central or the forward rapidity region. None of the models describes the full rapidity dependence.

The $\psi'$ cross section \cite{ypaper} is measured using the same data sample. In Fig.~\ref{fig:JPsi_y} (right) the weighted average computed from three decay channels, $\mu^+\mu^-\pi^+\pi^-, e^+e^-\pi^+\pi^-,$ $l^+l^-,$ is shown. The nuclear suppression factor for the $\psi'$ cross section for $x \in (0.3,1.6)\times 10^{-3}$ is computed to be $S_{\rm Pb} = 0.66\pm0.06$. This results is consistent with the value for $\rm{J/\psi}$.

Theoretical predictions which include shadowing (EPS09, LTA) agree with the measured cross section. Among the models involving saturation, the b-BK agrees with the observed data, while the GG-HS overpredicts it. The rest of the models overpredict the ALICE data point.

\subsection{$|t|$-differential cross section}

\begin{figure}[h!t]
\centering
\includegraphics[width=0.6\textwidth]{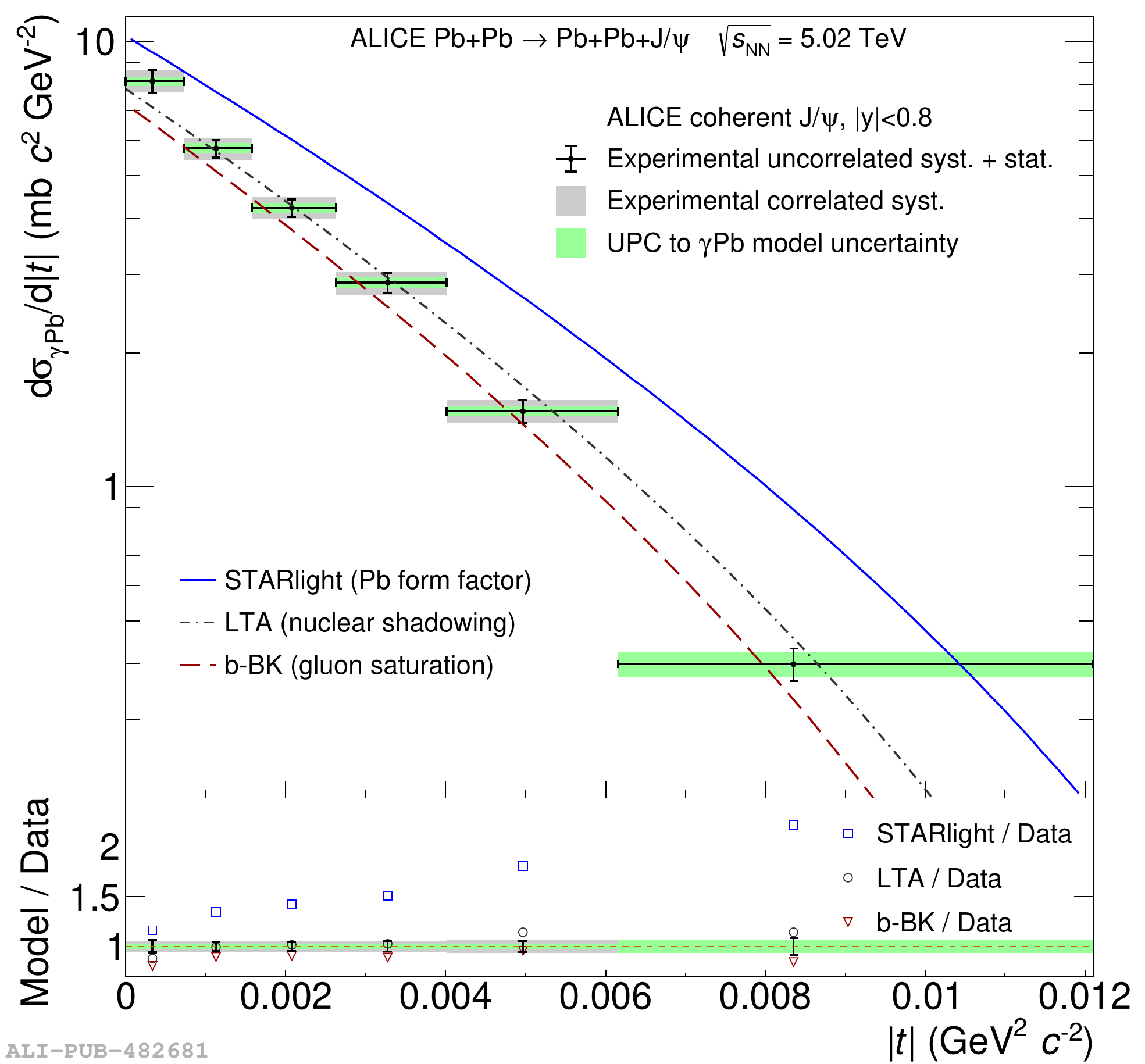}
\caption{$|t|$ dependence of the photonuclear cross section for the coherent $\rm{J/\psi}$ photoproduction compared with model predictions (top panel). Model to data ratio for each prediction in each measured point (bottom panel). The uncertainties are split between those originating from experiment and those originating from the correction to go from the UPC to the photonuclear cross section \cite{tpaper}.}
\label{fig:JPsi_t}
\end{figure}

The first measurement of the $|t|$-dependence of the coherent $\rm{J/\psi}$ photonuclear cross section~\cite{tpaper} was obtained at midrapidity from Pb–Pb UPCs at $\sqrt{s_{\rm NN}}=5.02$ TeV and it is shown in Fig.~\ref{fig:JPsi_t}. In this measurement only the dimuon decay channel was used as it has the smallest uncertainties. 

First, the $p^2_{\rm T}$-dependent UPC cross section was measured. To go from the measured $p^2_{\rm T}$~cross section distribution to the $|t|$-dependent UPC cross section, the $p^2_{\rm T}$ contribution of the photon was extracted using an unfolding technique. Further, the cross sections were corrected for the interference of the photon sources. Finally, to go from UPC to desired photonuclear cross section the photon fluxes were computed and used as follows, see e.g., \cite{guillermo} for details,
\begin{equation}
\left. \frac{{\rm d^2} \sigma^{\rm coh}_{\rm{J/\psi}}}{{\rm d} y {\rm d} p^2_{\rm T}}\right|_{y=0} = 2n_{\rm \gamma Pb} (y = 0) \frac{{\rm d} \sigma_{\rm \gamma Pb}}{{\rm d} |t|}.
\end{equation}

The difference between the ALICE results and the STARlight prediction (driven by the nuclear form factor) in both shape and magnitude demonstrates the presence of QCD dynamical effects in the $|t|$-dependence of the cross section. Models based on pQCD (LTA:nuclear shadowing, b-BK:gluon saturation) describe the data within the current uncertainties. Future measurements should allow us to distinguish between the two predictions.

\section{Conclusion and outlook}
The gluon content of nucleons in Pb nuclei at Bjorken-$x \in (0.3,1.4)\times 10^{-3}$ has been probed using ALICE data by measuring coherent charmonium photoproduction in Pb–Pb UPCs at $\sqrt{s_{\rm NN}}=5.02$ TeV. The nuclear suppression factor has been found to be $S_{\rm Pb} \approx 0.65$. The $|t|$~dependence of the cross section is observed to be sensitive to gluon distribution in the transverse plane.
Models incorporating shadowing using the LTA \cite{Frankfurt:2011cs} or saturation using the b-BK equation \cite{Bendova:2020hbb} describe all data at midrapidity within the current uncertainties. Results from the LHC might decrease the EPPS16 uncertainties as indicated in Ref. \cite{guzey} where a reweighing of the EPPS16 nPDFs was implemented.

In the Run 3 and 4 of the LHC the integrated luminosity will increase by a factor 13 with respect to the previous Run 2. Moreover, ALICE will run in continuous readout mode with higher data collection efficiency. There are also significant detector upgrades. All of this will result in millions of $\rm{J/\psi}$ to be recorded by ALICE \cite{outlook}. Thus a larger data sample, smaller systematic uncertainties and a better efficiency will lead to increased precision on all previous measurements as well as enabling to make new differential measurements, e.g., ${\rm d^2} \sigma/{\rm d} y {\rm d} |t|$, and angular dependencies between the decay products. In addition, completely new measurements will become possible, e.g., photoproduction of $\Upsilon$ \cite{outlook}, which occurs at a 10 times larger scale that the $\rm{J/\psi}$, and measurement of interference effects \cite{Klein:1999gv} between the two source-target configurations of the heavy ions.



\paragraph{Funding information}

This work has been partially supported by the grant 18-07880S of the Czech Science Foundation (GACR).

\bibliography{ms.bib}

\nolinenumbers

\end{document}